\documentstyle[sprocl,psfig]{article}

\def\be{\begin{equation}}
\def\ee{\end{equation}}
\def\bea{\begin{eqnarray}}
\def\eea{\end{eqnarray}}

\begin{document}

\title{VALIDITY OF FLAVOR SYMMETRY AND CHARGE SYMMETRY FOR PARTON
DISTRIBUTIONS}

\author{J.T. Londergan}

\address{Dept. of Physics and Nuclear Theory
	Center, Indiana University \\ 
        Bloomington, IN 47405, USA;  \\E-mail: tlonderg@iucf.indiana.edu
	}


\maketitle\abstracts{ Recent experimental measurements of the 
Gottfried Sum Rule, and pp and pD Drell-Yan processes,  
suggest significant violation of flavor symmetry in the proton sea. 
This interpretation rests 
on the assumption of parton charge symmetry. Our model
calculations suggest charge symmetry violation [CSV] for parton
valence distributions of a few percent. Precision measurements 
of structure functions in muon and neutrino reactions allow us to set 
rather stringent experimental limits on CSV in certain kinematic
regions.  In another region, these experiments suggest substantial  
CSV effects. We suggest  
experiments which could test parton CSV.}

\section{Flavor Symmetry in Parton Distributions}
The basic features of parton distributions have been well
established through measurements of deep inelastic
scattering [DIS], Drell-Yan processes and direct photon experiments.  
Precision tests of approximate symmetries allow us to understand
the details of nucleon parton distributions.  For example, we
know that the strange quark distribution is substantially smaller
than the light nonstrange sea, due to the relatively large
mass of the $s$ quark (this is sometimes termed SU(3) flavor
symmetry violation) \cite{Baz95}.  
A new generation of precise high energy experiments allows
us to examine finer details of parton distributions.  An example
of this is the NMC experiment \cite{NMC}, which measured 
$\mu p$ and $\mu D$ DIS, and accurately tested the Gottfried Sum 
Rule $S_G$ by comparing $F_2^{\mu p}$ and $F_2^{\mu n}$.  
Assuming $\bar{d}^p(x) = \bar{u}^p(x)$ one predicts  
$S_G = 1/3$. pQCD predicts very small deviations from 1/3. The NMC 
result $S_G = 0.235 \pm 0.026$ 
was four standard deviations lower than
the ``naive'' prediction, apparently indicating 
significant flavor symmetry violation [FSV] in the proton sea. 

This was followed by a comparison of 
$pp$ and $pD$ Drell-Yan [DY] processes \cite{E866}. 
For large $x_F$ the ratio of DY cross sections 
will be larger than one if $\bar{d}^p(x) > \bar{u}^p(x)$, as 
observed in the E866 experiment (for a detailed discussion 
see the talk by W. Melnitchouk
at this conference). The most promising theoretical 
model to date is the ``meson-cloud'' picture.
In these models one includes a quark ``core'' for the nucleon plus 
a ``cloud'' of baryon-meson Fock components, and   
the virtual photon scatters from any of these 
components.  Melnitchouk showed that quantitative agreement with E866 
data can be achieved with a model 
including nucleon, pion and $\Delta$ components.  

\section{Charge Symmetry Violation in Parton Distributions}
At first sight, the DY and NMC experiments appear to show 
a large FSV contribution to the proton sea.  
However, {\em all these results
depend on the assumption of parton charge symmetry}.  Ma \cite{Ma92} 
showed that both the DY and NMC experiments could be reproduced, even 
if flavor symmetry was exact, by assuming a sufficiently large violation
of parton charge symmetry. In this talk we examine the following
questions: 1) Are there theoretical grounds for expecting parton CSV?
2) What are the present experimental limits on parton 
charge symmetry? 3) What are the most promising experiments which could
improve the current limits on parton CSV?

\subsection{A Model for Parton Charge Symmetry Violation}
Charge symmetry for parton distributions has been investigated
recently by several groups \cite{Sat92,Ben98}.  We 
review here the work of Benesh and Londergan \cite{Ben98}.  This is 
based on the Adelaide model 
for evaluating twist-two parton distributions with proper support.  
It involves evaluating contributions to parton distributions through
the relation 
\begin{equation}
q(x,\mu^2) = M\sum_X \,|\langle X|\psi_+(0)|N\rangle |^2 \,\delta(M(1-x)
  -p_X^+)
\label{adel}
\end{equation}
In Eq.\ \ref{adel}, $\psi_+ = (1+\alpha_3)\psi/2$, and $X$ 
represents a complete set of eigenstates of the Hamiltonian 
$H$.  The parton distribution $q(x,\mu^2)$ is guaranteed to
have proper support, i.e.\ it vanishes by construction for
$x>1$.  

\begin{figure}[t]
\centering{ \hbox{ \hspace{0.3cm} 
\psfig{figure=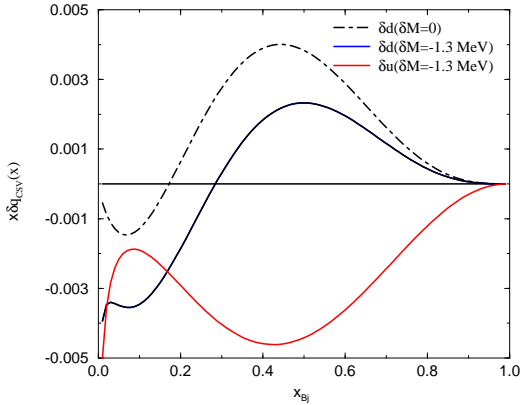,height=2.4in} \hspace{0.3cm}
  }}
\caption{CSV valence quark distributions $x\delta u(x)$ (neg) 
and $x\delta d(x)$ (pos), taken from Ref.\ \protect\cite{Ben98}. CSV 
terms arise from mass difference in spectator diquark, and from 
$n-p$ mass difference.}  
\label{fig1}
\end{figure}

For relatively large $x$ values, the dominant contribution to the
valence quark distribution comes from the lowest  
two-quark spectator state contributing to Eq.\ \ref{adel}. In 
this case, we can derive an analytic form for the change in the
quark distribution $\delta q(x)$ arising from a small change 
$\delta m$ in the diquark mass, 
\begin{equation}
  \delta q(x) \approx {2m\,\delta m (1-x) \over M^2(1-x)^2 + m^2} 
  {d q(x) \over dx}  
\label{analyt}
\end{equation}
From this equation, we can estimate the magnitude of charge symmetry 
violation directly from phenomenological parton distributions without
using quark models.  The results obtained are in very good agreement
with direct quark model calculations of CSV effects through Eq.\ 
\ref{adel}.  Alternatively, we can relate CSV effects to 
spin-flavor effects on parton distributions.  In Fig.\ \ref{fig1}, we 
show calculations of parton charge symmetry violation in which
we include both diquark mass contributions and nucleon mass
differences; the quark CSV terms are defined through  
\begin{eqnarray}
 \delta d_{\rm v}(x) &=& d^p_{\rm v}(x)- u^n_{\rm v}(x) \nonumber \\ 
 \delta u_{\rm v}(x) &=& u^p_{\rm v}(x)- d^n_{\rm v}(x)
\label{csvdef}
\end{eqnarray}

The theoretical parton CSV terms are predicted to be approximately
equal and opposite, i.e.\ $\delta u_v(x) \approx - \delta d_v(x)$.   
Since at large $x$ we have $d_v(x) << u_v(x)$, the fractional 
CSV term will be much larger for the ``minority quark'' term $d_v(x)$ 
than for $u_v(x)$.  We predict $\delta d_v(x)/d_v(x)$ to be of order 
3-6\% at large $x$.   
The effect shown here is sufficiently large that one could
question its reliability.  However, it appears to be robust since 
it is obtained through either simple quark models, or through the 
analytic result using, e.g., CTEQ parton distributions\cite{CTEQ}.  It 
is important that this prediction be verified 
experimentally.  However, it requires experiments which 
specifically probe the ``minority'' quark distribution, since this
is substantially smaller than the ``majority'' quark distribution
at large $x$.   

\subsection{Experimental Status of Parton Charge Symmetry}
The most sensitive experimental test of parton charge
symmetry to date is the ``charge ratio''.  There
is a simple relation between the $F_2$ structure functions for 
charged lepton DIS and neutrino charged current reactions on an 
isoscalar target $N_0$: 
\begin{eqnarray}
R_c(x) &=& {F_2^{\gamma N_0}(x) \over {5\over 18} F_2^{\nu N_0}(x)
  - {x (s(x)+\bar{s}(x))\over 6} } \approx 1 + 
  {\bar{s}(x) -s(x) \over \bar{Q}(x)} 
  \nonumber \\ &+&
   { 4(\delta u_v(x) -\delta d_v(x)) + \delta \bar{u}(x) -\delta 
   \bar{d}(x) \over 5 \bar{Q}(x) } \nonumber \\ 
   \bar{Q}(x) &=& \sum_{j=u,d,s} q_j(x) + \bar{q}_j(x) - {3x(s(x) + 
   \bar{s}(x)) \over 5}
\label{chrat}
\end{eqnarray}
The relation $R_c(x) =1$ should
hold for all $x$ and $Q^2$, with no QCD corrections.  
The only things which break this relation are parton CSV terms, or 
contributions from $s(x) \ne \bar{s}(x)$. 

\begin{figure}[t]
\centering{ \hbox{ \hspace{0.3cm} 
\psfig{figure=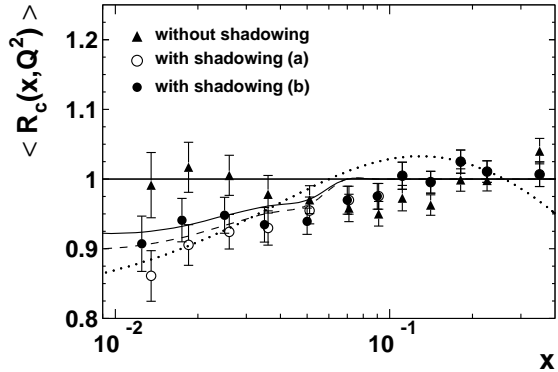,height=2.3in} \hspace{0.3cm}
 }}
\caption{Charge ratio $R_c(x)$ of Eq.\ \protect\ref{chrat}, obtained 
using NMC $\mu$-D structure functions and CCFR $\nu$-iron. Solid circles: 
heavy target shadowing corrections calculated specifically for 
neutrinos; open circles: $\nu$ heavy target corrections taken from 
shadowing observed in charged lepton DIS.}  
\label{fig2}
\end{figure}

In Fig.\ \ref{fig2} we plot $R_c(x)$, 
the ratio of NMC $\mu$-D structure functions \cite{NMC} to CCFR 
$\nu$-Fe measurements \cite{CCFR}. 
The solid circles show the ratio $R_c(x)$, when the heavy target
corrections are calculated specifically for neutrinos.  For 
intermediate values 
$x > 0.1$ the agreement between structure functions is very good, and 
we can set upper limits of a few percent on parton CSV.
However, in the region $x < 0.1$, $R_c$ deviates from unity 
by as much as 10\%.  This is discussed in the
paper by C. Boros at this conference.  From Eq.\ \ref{chrat} it
would appear that the low-$x$ discrepancy could be accommodated
by allowing $s(x) \ne \bar{s}(x)$.  However, if one combines 
NMC and CCFR data with opposite-sign dimuon production data from
neutrino reactions (which is used to extract $s(x)$), then one 
can show that the discrepancy cannot be removed unless one takes 
$\bar{s}(x) < 0$, which is not physically reasonable \cite{Bor98}.  
Thus, if the existing data are correct, they are not compatible unless 
one assumes a very large sea quark CSV effect (roughly 25\%) at small $x$.  

\section{Proposed Experimental Tests of Parton Charge
Symmetry}

\subsection{Test of Weak Current Relation $F_1^{W^+ N_0}(x) = 
F_1^{W^- N_0}(x)$}
At sufficiently high energies, the charge-changing structure
functions on an isoscalar target are equal except for
contributions from valence quark CSV, and possible strange
or charmed quark terms, i.e.\ 
\begin{eqnarray}
 { 2(F_1^{W^+ N_0}(x,Q^2) - F_1^{W^- N_0}(x,Q^2) )\over 
  F_1^{W^+ N_0}(x,Q^2) + F_1^{W^- N_0}(x,Q^2) } &\approx&  
  { \delta d_{\rm v}(x) - \delta u_{\rm v}(x) \over Q(x) }
  \nonumber \\ &+& { 2( s(x) - \bar{s}(x)) \over Q(x)} , \nonumber \\ 
  &\equiv& R_{\scriptscriptstyle{CSV}}(x) + R_s(x) \nonumber \\ 
  Q(x) &=& \sum_{j=u,d,s} q_j^p(x) + \bar{q}^p_j(x) ~~. 
\label{hera}
\end{eqnarray}
In Eq.\ \ref{hera} we have expanded to lowest order in the small 
CSV terms. 
At the enormous values of $Q^2$ that can be probed at HERA, weak
interaction processes such as $e^- p \rightarrow \nu_e X$
are not impossibly small compared to the electromagnetic process
$e^- p \rightarrow e^- X$.  If deuteron beams were available at
HERA, this would provide a very clean test of charge symmetry, and/or
the equality of strange/antistrange quark distributions. 
The $(e^-,\nu_e)$
reaction picks out positively charged partons in the target, 
while the $(e^+,\bar{\nu}_e)$ 
reaction measures the negatively charged partons.  

\begin{figure}[t]
\centering{ \hbox{ \hspace{-0.15cm} 
\psfig{figure=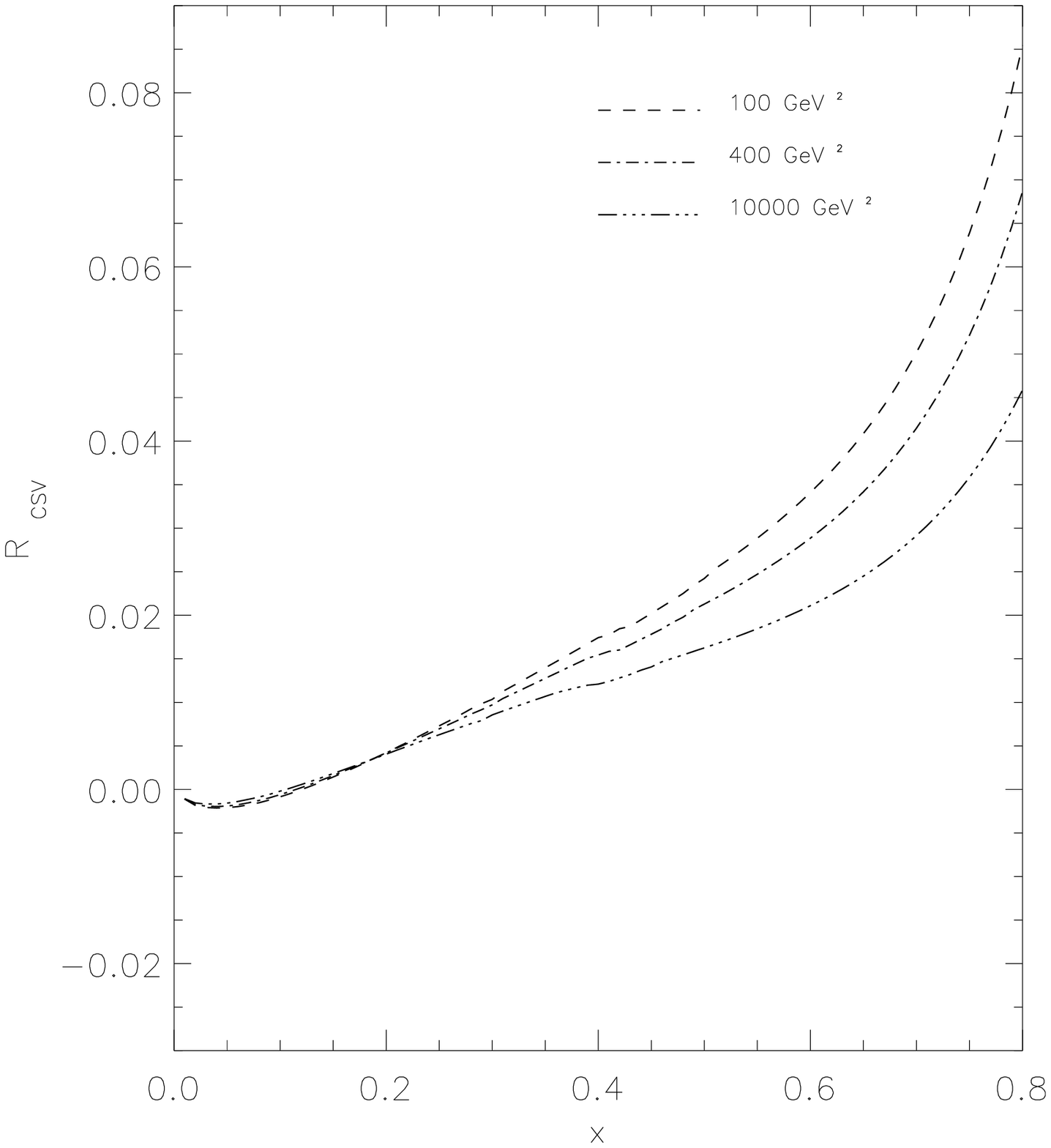,height=2.4in} \hspace{0.4cm} 
\psfig{figure=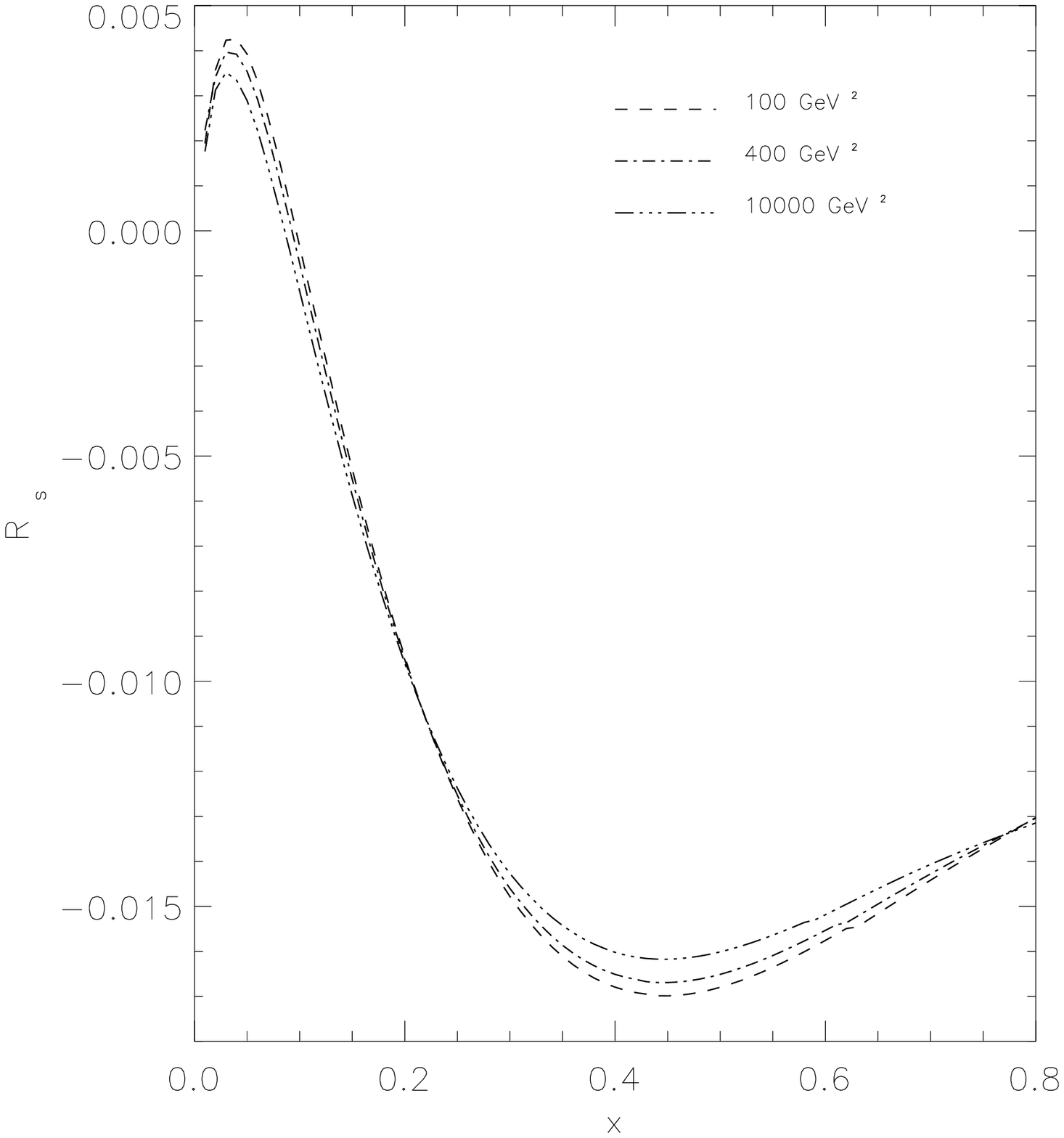,height=2.4in} 
 }}
\caption{(L) The charge symmetry violating ratio 
$R_{\scriptscriptstyle{CSV}}$ defined in Eq.\ \protect\ref{hera}. 
(R) The contribution to the difference in $e^\pm$ charge changing 
processes arising from a difference $s-\bar{s}$ labeled $R_s$ in 
Eq.\ \protect\ref{hera}.}  
\label{fig3}
\end{figure}

The difference between the structure functions $F_1^{W^+D}$ and 
$F_1^{W^-D}$ has been studied recently \cite{Lon97}.  The results
are shown in Fig.\ \ref{fig3}; the $s-\bar{s}$ term was taken from the 
model of Melnitchouk and Malheiro\cite{Mel97}.  For sufficiently 
large $x$ values predicted results are as large as a few percent.   
This experiment would provide a strong test of
charge symmetry in parton distributions, and would require almost none
of the corrections necessary for the ``charge ratio'' test.  
Note that at small $x$ this comparison could 
test whether $\bar{s}(x) = s(x)$, and thus distinguish between the 
strange/antistrange and CSV contributions to the small-$x$ charge ratio
discrepancy.  

\subsection{Drell-Yan Processes Initiated by Charged Pions
on Isoscalar Targets}
One way to test parton charge symmetry is to compare 
DY processes for charged pions on isoscalar targets.  
This is most evident in the ``valence-dominated'' region, 
where both $x_\pi$ and $x$ are large.  In this region, 
\begin{equation} 
  |\pi^+\rangle \sim u^{\pi^+}\bar{d}^{\pi^+}~~; \quad 
  |\pi^-\rangle \sim d^{\pi^+}\bar{u}^{\pi^+}~~; 
  \quad  |p(n)\rangle \sim u_{\rm v} 
  + d_{\rm v} \quad .
\end{equation}
Thus, in $|\pi^+\rangle$ ($|\pi^-\rangle$) DY, a 
$\bar{d}$ ($\bar{u}$) in the pion 
will annihilate a down (up) valence quark in the nucleon. We can
test charge symmetry by forming the following ratio for an 
isoscalar target $N_0$:
\begin{equation}
 R^{DY}_{\pi N_0}(x,x_\pi) =  \frac{4 \sigma_{\pi^+N_0}^{DY}(x,x_\pi) - 
  \sigma_{\pi^-N_0}^{DY}(x,x_\pi)}
  {\left(4 \sigma_{\pi^+N_0}^{DY}(x,x_\pi) + 
  \sigma_{\pi^-N_0}^{DY}(x,x_\pi) \right) /2} \approx  
  \left( \frac {\delta d(x) - \delta u(x)}{u^p_v(x) + d^p_v(x)} 
  \right)
\label{eq:R}
\end{equation}
This has been investigated by Londergan et al.\cite{Lon94}, who   
conclude that CSV could be tested even in the presence
of valence-sea interference terms (not shown in Eq.\ \ref{eq:R}).  
With sufficiently intense pion beams, these 
measurements could decrease the current upper limits on quark CSV.  

\subsection{Charged Pion Leptoproduction from Isoscalar Targets}
A process like $e^- + A \rightarrow \pi^{+ (-)} + X$ 
could also be a sensitive probe of CSV in nucleon  
valence distributions.  There are ``favored''
and ``unfavored'' fragmentation modes; for example, a $u$ quark
is more likely to fragment into a $\pi^+$ which contains a $u$ valence
quark.  Tests of CSV involves comparison  of $\pi^+$ and $\pi^-$ 
electroproduction from isoscalar targets. Londergan, Pang and
Thomas \cite{Lon96} concluded that CSV tests would be feasible in 
this process.  This quantity could in principle be obtained at 
the HERMES experiment 
at HERA, which is presently measuring pion fragmentation
functions for electroproduction on deuterons. 

\section{Conclusions}
Suggestions of large SU(2) FSV, i.e. $\bar{d}^p(x) > \bar{u}^p(x)$,   
are confirmed by FNAL experiment E866,
which compared $pp$ and $pD$ DY processes. 
pQCD contributions are too small for experiment, but  ``meson-cloud'' 
models achieve quantitative success.  
The conclusion that flavor symmetry is broken rests on the
implicit assumption of parton charge symmetry.
Theoretical calculations suggest valence quark CSV of a few percent. 
Current experiments set upper 
limits of a few percent on parton CSV for $x > 0.1$, but suggest 
uncomfortably large CSV effects for $x < 0.1$. We suggest three 
experiments which could accurately test parton CSV.  The first 
compares structure functions measured in weak charge-changing 
reactions which could be accessed in $e^- +D$ and $e^+ +D$ reactions
at HERA.  The second type of reaction compares DY processes for
$\pi^+$ and $\pi^-$ beams on isoscalar targets; this might be done
in fixed-targets at FNAL following the Main Injector upgrade.  
The third reaction compares charged pion leptoproduction from 
isoscalar targets; this experiment is currently being carried 
out in the HERMES experiment at HERA. 

\section*{Acknowledgments}
This research was performed in collaboration with A.W. Thomas.  
The author would also like to thank other collaborators, 
C. Benesh, C. Boros, S. Braendler, G. Liu, A. Pang and E. Rodionov. 
He also thanks the Special Research Centre for the Subatomic Structure 
of Matter for its hospitality.  This research was supported in part by 
the NSF under contract NSF-PHY/9722706, and by the Australian
Research Council.


\section*{References}
\begin{thebibliography}{99}

\bibitem{Baz95} A.O. Bazarko et al., Z. Phys. {\bf 65}, 189 (1995).
\bibitem{NMC} P. Amaudruz et al. (NMC Collaboration,) Phys.Rev.
	Lett. {\bf 66}, 2712 (1991).
\bibitem{E866} E.A. Hawker et al. (E866 Collaboration), Phys. Rev. 
	Lett. {\bf 80}, 3715 (1998).
\bibitem{Ma92} B.-Q. Ma, Phys.Lett. {\bf B274}, 111 (1992).
\bibitem{Sat92} E. Sather, Phys.Lett. {\bf B274}, 433 (1992); 
	E. Rodionov, A.W. Thomas and J.T. Londergan, 
	Int.J.Mod.Phys.Lett. {\bf A9}, 1799 (1994); C.J. Benesh and 
	T. Goldman, Phys.\ Rev.\ {\bf C55}, 441 (1997).
\bibitem{Ben98} C.J. Benesh and J.T. Londergan,  Phys.Rev. C, 
	to be published, 1998 (preprint nucl-th/9803017).  
\bibitem{CTEQ} H.L. Lai {\it et al.}, Phys. Rev. {\bf D55}, 1280 
	(1997).
\bibitem{CCFR} W.G. Seligman et al. (CCFR Collaboration), Phys.\ 
	Rev.\ Lett.\ {\bf 79}, 1213 (1997).
\bibitem{Bor98} C. Boros, J.T. Londergan and A.W. Thomas,  
	to be published (preprint hep-ph/9806249).
\bibitem{Lon97} J.T. Londergan, S.Braendler and A.W. Thomas,  
	Phys.\ Lett.\ {\bf B424}, 185 (1998).
\bibitem{Mel97} W. Melnitchouk and M. Malheiro, Phys. Rev. {\bf C55}, 
	431 (1997). 
\bibitem{Lon94} J.T. Londergan, G.T. Garvey, G.Q. Liu, 
	E.N. Rodionov and A.W. Thomas, Phys.Lett. {\bf B340}, 115 (1994).
\bibitem{Lon96} J.T. Londergan, Alex Pang and A.W. Thomas, Phys.Rev. 
	{\bf D54}, 3154 (1996).  
	
	
\end{thebibliography}
\end{document}